\documentclass[12pt,a4paper]{article}
\usepackage{epsf,pifont}

\begin{document}

\baselineskip 20pt     

\begin{center}

{\bf
BIMODALITY IN SPECTATOR FRAGMENTATION
}
\vspace*{0.6cm}

\normalsize                

W. Trautmann$^{(a)}$ and the ALADIN Collaboration\\
\vspace*{0.3cm}
   {\small \it  $^{(a)}$ Gesellschaft f{\"u}r Schwerionenforschung (GSI),\\
Planckstr. 1, D-64291 Darmstadt, Germany }\\

\end{center}

\begin{abstract}
The fluctuations of the largest fragment charge of a partition 
and of the charge asymmetries of the two or three largest fragments in 
spectator decays following 
$^{197}$Au + $^{197}$Au collisions at 1000 MeV per nucleon 
are investigated. The observed bimodal distributions at specific 
values of the sorting variable $Z_{\rm bound}$ exhibit features known 
from percolation theory where they appear as finite-size effects.
The underlying configurational fluctuations seem generic for
fragmentation processes in small systems. 

\end{abstract}

\normalsize                

\section{Introduction}
\label{sec:intro}

Double-humped event distributions have received particular interest 
recently because their observation might indicate bimodality which 
is one of the signals expected for a first-order phase transition 
in finite systems 
[1-3]. Bimodality occurs when non-negligible surface 
interactions at the phase boundary lead to a convex entropy function 
in the transition region. Canonical sampling near the transition 
temperature will then produce two distinct event classes which 
differ with respect to the order parameter of the transition. 

Bimodality has, e.g., been observed in solid-liquid transitions of 
clusters of Na atoms \cite{schmidt01}. In these experiments, the clusters 
were thermalized in a heat bath of helium gas and excited with photons from
a laser beam.
Corresponding experiments in nuclear fragmentation face the difficulty 
that the temperature cannot be predetermined and that
a canonical sampling can thus not be performed. 
There is, furthermore, the possibility of impact-parameter mixing, 
meaning that fluctuations of the variables used for sorting 
will unavoidably lead to finite distributions with respect to other 
observables or event characteristics (e.g., impact parameter) 
even in narrowly selected event samples.
It is, nevertheless, of interest to study the origin and the meaning of
bimodal event distributions and the conditions under which they occur.
For example, Lopez et al. using the HIPSE event generator have recently 
pointed to the role of angular momentum in producing bimodal 
distributions as a result of instabilities of nuclei with high spin 
\cite{lopez05}. Pichon et al. \cite{pichon06} 
have shown that the two bumps of the bimodal distributions observed
in $^{197}$Au + $^{197}$Au fragmentations at 60 to 100~MeV per nucleon
correspond to different scaling properties of the distributions of the 
largest fragment charge ($\Delta$-scaling \cite{botet01}). 

In this work, the fluctuations of the largest fragment charge of a 
partition and of the charge asymmetries of the two or three largest 
fragments from the decay of excited projectile spectators in $^{197}$Au + 
$^{197}$Au collisions at 1000 MeV per nucleon are investigated.
The data have been collected in experiments performed 
with the ALADIN spectrometer at GSI.
The observed bimodal distributions at specific values of the 
sorting variable $Z_{\rm bound}$ exhibit features known from percolation 
theory where they appear as finite-size effects. Percolation on a large
lattice exhibits signatures of a second-order phase 
transition \cite{stauffer}.  
The observed similarities thus raise the question whether 
bimodality in fragmentation reactions may be used to infer the
order of the phase transition in the nuclear case.

Bimodality and its significance for the interpretation of fragmentation 
data and their relation to the nuclear liquid-gas phase transition have 
been frequently discussed at previous conferences of this series 
[8-12]. At this 
year's conference, new results obtained by the INDRA collaboration are 
presented by E. Bonnet \cite{bonnet07}.

\begin{figure}
\begin{minipage}[t]{.47\textwidth}	
    \centering
    \epsfysize=6.7cm

   \epsffile{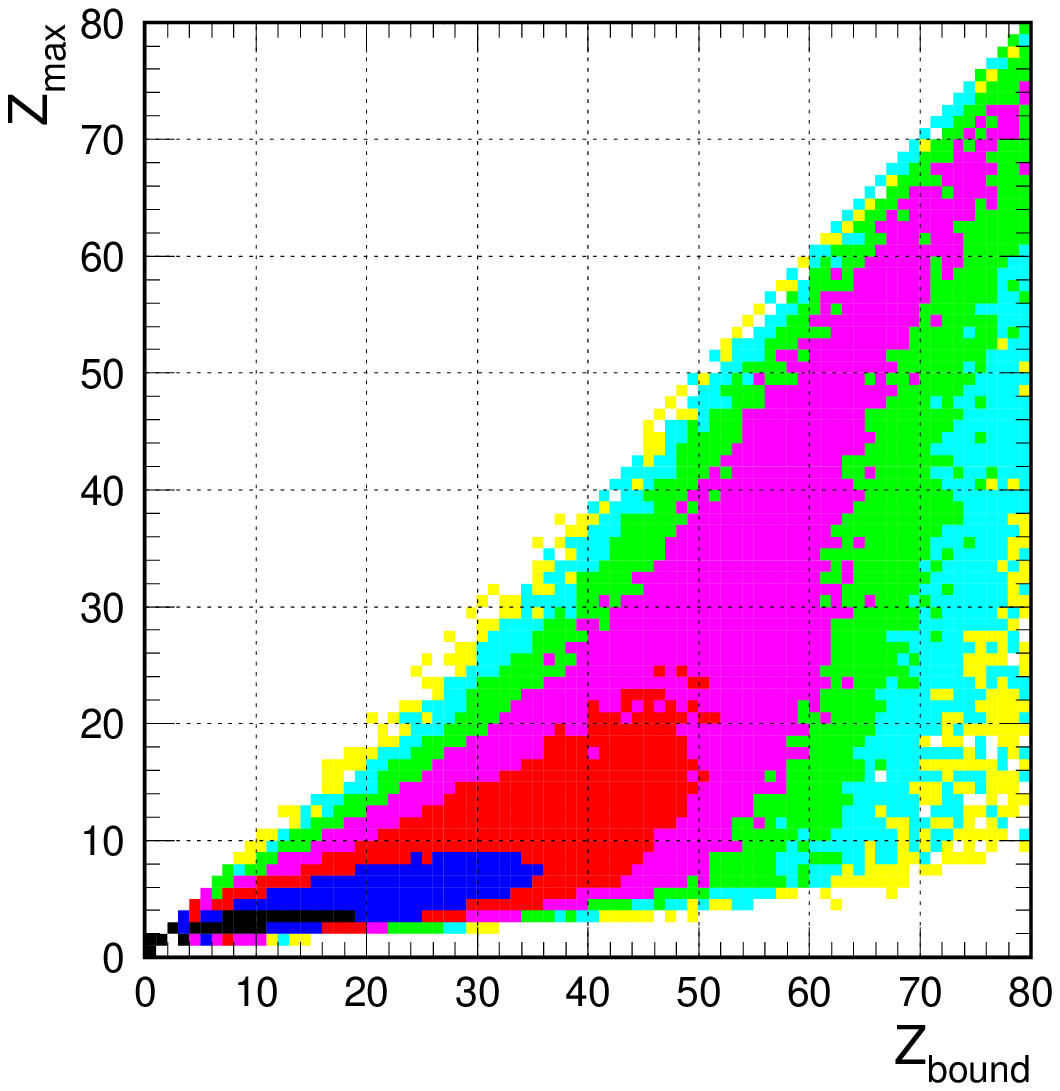}
\vspace{-4mm}

\caption{Distribution of $Z_{\rm max}$ versus $Z_{\rm bound}$ for 
projectile fragments from $^{197}$Au on 
$^{197}$Au at 1000 MeV per nucleon \protect\cite{schuett96}. 
Conventional fission events are removed. The shadings follow a 
logarithmic scale.
}

\label{fig:zbound}
\end{minipage}
\hspace{\fill}
\begin{minipage}[t]{.47\textwidth}	

    \centering
    \epsfysize=6.5cm

   \epsffile{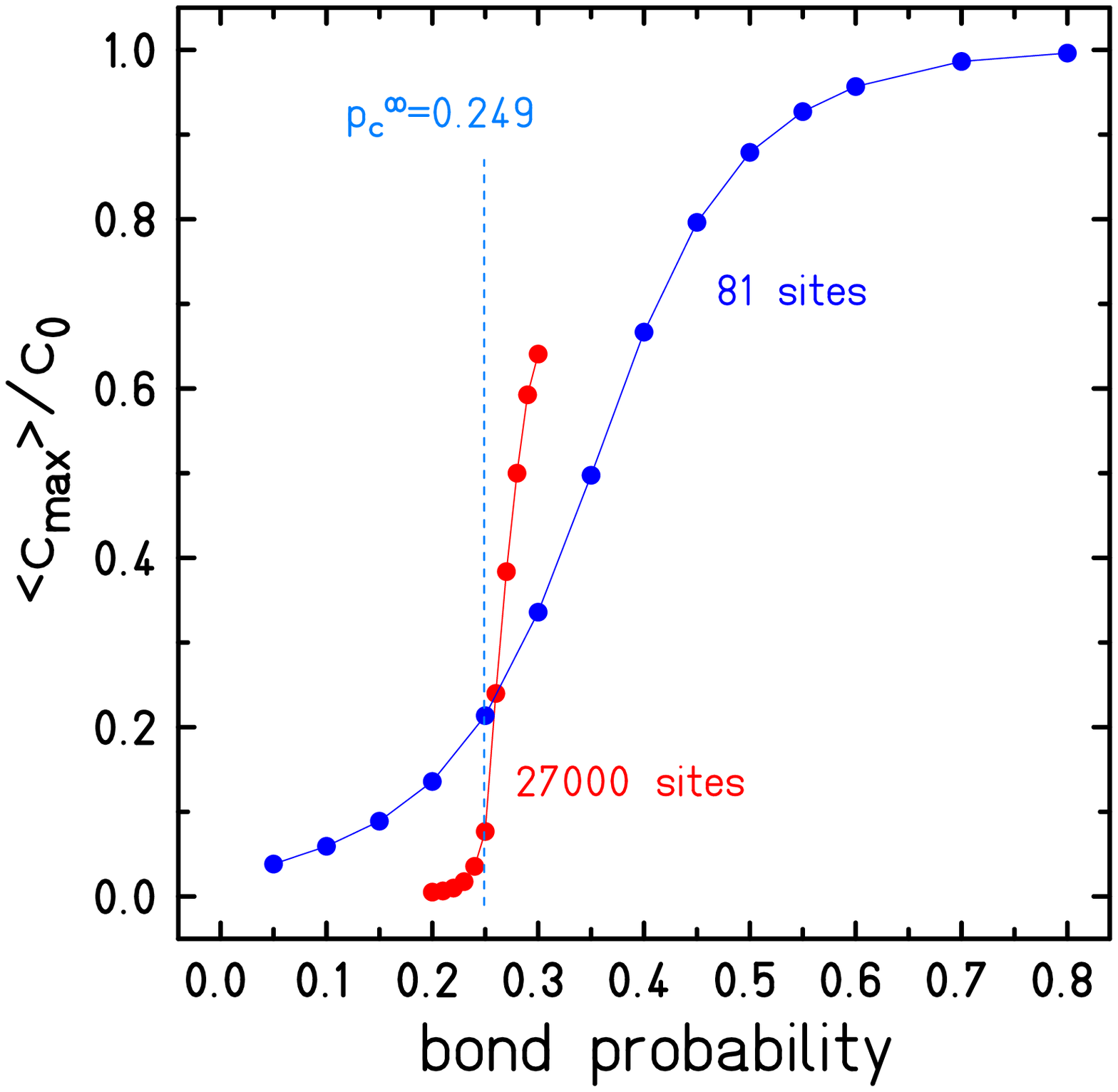}

\vspace{-4mm}

\caption{Bond percolation: mean relative magnitude of the largest cluster 
as a function of the bond probability for cubic lattices of $c_0$~=~81 and 
27000 sites. 
The critical bond probability in the infinite sytem is 
$p_{c}^{\infty}$~=~0.249. 
}

\label{fig:perco1}
\end{minipage}
\end{figure}

\section{Experimental results}
\label{sec:exper}

The data used for the present analysis were obtained by the ALADIN 
collaboration in measurements using $^{197}$Au projectiles of 1000 MeV per 
nucleon delivered by the heavy-ion synchrotron SIS at 
GSI \cite{poch95,schuett96}. The ALADIN spectrometer was used to detect 
and identify the products of the projectile-spectator decay following 
collisions with $^{197}$Au target nuclei. 

The sorting variable
$Z_{\rm bound}$ is defined as the sum of the atomic numbers $Z_i$ of 
all projectile fragments with $Z_i \geq$ 2. 
It reflects the variation of the 
charge of the primary spectator system and is monotonically
correlated with the impact parameter of the reaction \cite{ogilvie93}. 
The evolution of the dominant reaction processes is illustrated
in Fig.~\ref{fig:zbound} which shows the 
correlation of the largest atomic number $Z_{\rm max}$ observed in a 
partition with $Z_{\rm bound}$.
Large values of $Z_{\rm bound}$ correspond to low 
excitation energies, at which the
decay changes its character from evaporation-like processes 
($Z_{\rm max}~\approx~Z_{\rm bound}$) to multifragmentation 
("rise" of multifragmentation) while 
small values correspond to reaction channels
with high excitation energies and disintegrations into predominantly very
light clusters ("fall" of multifragmentation, 
$Z_{\rm max} \ll Z_{\rm bound}$). 

\begin{figure} [!htb]	
    \leavevmode
    \centering
    \epsfxsize=0.65\linewidth

   \epsffile{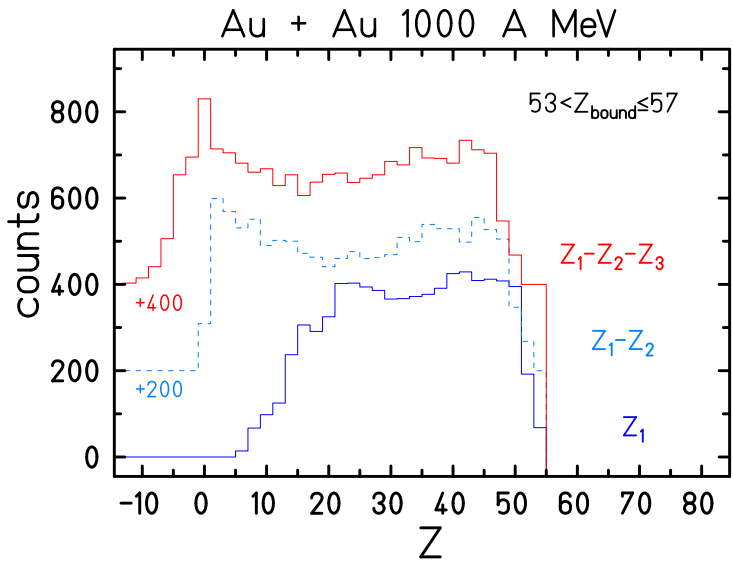}


\caption{{Distributions of the atomic number $Z_1$ of the largest fragment
of a partition and of the differences $Z_1-Z_2$ and $Z_1-Z_2-Z_3$ for events with
$53 < Z_{\rm bound} \leq 57$ from the fragmentation of $^{197}$Au projectiles 
at 1000 MeV per nucleon. 
Note the offsets by 200 and 400 counts of the difference distributions.}}

\label{fig:spect} \end{figure}

Besides the evolution of the mean and of the fluctuations of $Z_{\rm max}$
(alternatively denoted by $Z_1$ in the following) 
also those of two-fragment and three-fragment asymmetries are of
interest and characterize the dominant transition of the reaction 
mechanism \cite{kreutz93}. It is found that the ratios 
$\langle Z_2 /Z_1 \rangle$ and $\langle Z_3 /Z_2 \rangle$ both approach 
$\approx 0.6$ at small $Z_{\rm bound}$ 
($Z_2$ and $Z_3$ are the second and third largest atomic number 
$Z$ of a partition). 
Consequently, the charge difference $Z_1-Z_2-Z_3$, or 
the corresponding asymmetry after normalizing with respect to the system 
charge $Z_0$, will approach zero at small $Z_{\rm bound}$ while it is 
close to $Z_0$ (the asymmetry close to 1) at large $Z_{\rm bound}$. 
In the transition region, the fluctuations of these observables are large
\cite{kreutz93} and the distributions are bimodal, i.e. they exhibit a
two-hump structure (Fig.~\ref{fig:spect}). Note that 
$\langle Z_{\rm max} \rangle$ 
drops most rapidly in the bin $53<Z_{\rm bound}\le 57$
at which this bimodality is most strongly pronounced 
(Fig.~\ref{fig:zbound}).

\section{Largest fragment as order parameter}
\label{sec:largf}

In the search for an experimentally accessible order parameter 
of the nuclear liquid-gas phase transition, as observed in multifragmentation
reactions, the magnitude of the largest 
fragment of the partition has appeared as a promising choice. 
It may be identified with the part of the system in the liquid phase, 
and it is correlated with the mean density which is the natural 
order parameter of a liquid-gas phase transition. Observables correlated
with it, as e.g. the differences and asymmetries discussed above,
may similarly serve as order parameters. 

Statistical model calculations for nuclear multifragmentation
show that the disappearance of the dominating fragment is
associated with a maximum of the heat capacity which is the more 
strongly pronounced the larger the system \cite{dasgupta98}. 
For $A_0$~=~150, the system mass for $Z_{\rm bound}\approx$ 50
\cite{poch95}, the predicted specific heat distribution is rather wide
with a maximum at $T \approx 6.3$~MeV. This transition temperature, or 
boiling temperature according to the authors of \cite{dasgupta98}, 
is comparable with values
of the double-isotope temperature $T_{\rm HeLi}$ as measured for 
the present reaction \cite{poch95,xi97} and for similar systems
\cite{kelic06}. 
The good description of the charge correlations and charge asymmetries 
characterizing the partitioning of the system, including their variances, 
with statistical multifragmentation models provides further evidence 
for the first-order nature of the transition \cite{xi97,botv95}. 
Bimodality is predicted for canonical ensembles \cite{buyuk05,chaudhuri07}.

The disappearance of the largest cluster, with the variation of a 
suitable control parameter, has been identified as a prominent 
signal also in fragmentations of other systems as, e.g., 
atomic hydrogen clusters \cite{gobet01}, and the extension of the 
largest cluster 
is an order parameter in percolation theory \cite{stauffer}. 
On finite percolation lattices, the disappearance of 
a dominant largest cluster proceeds rather smoothly and with 
obvious similarity to the nuclear experiment 
(Figs.~\ref{fig:zbound},\ref{fig:perco1}).

\begin{figure} [!htb]	
    \leavevmode
    \centering
    \epsfysize=11.0cm

   \epsffile{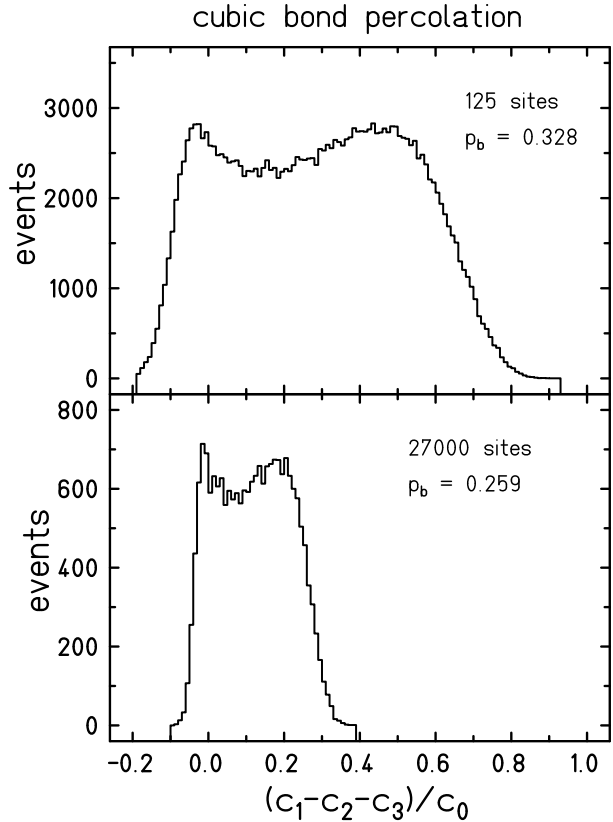}


\caption{{Examples of the distributions of the 3-fragment difference
$c_1-c_2-c_3$, normalized with respect to the lattice size $c_0 = L^3$, as 
obtained with cubic bond percolation for the cases $c_0 = 125$ sites and
$p_b=0.328$ (top) and $c_0 = 27000$ sites and $p_b=0.259$ (bottom).}}

\label{fig:perco2} \end{figure}

\section{Cubic bond percolation}
\label{sec:perco}

Percolation models have been quite successfully used for
describing the properties of nuclear fragmentation 
[17,24-28] including the apparent 
critical behaviour. For the present purpose, calculations with a 
cubic-bond-percolation model have been performed with various lattice 
sizes. The critical bond parameter for this type of lattice is
$p_{c}^{\infty}$~= 0.249 \cite{stauffer,pbinf}. For large lattices, in the
limit of infinity, a sharp transition with the sudden appearance 
of an extended percolating cluster is observed for this value of the 
probability that a bond exists between neighbouring sites.
For finite
lattices, the transition is smooth (Fig.~\ref{fig:perco1}) and, for a
lattice of 81 sites (obtained by smoothing the corners and edges 
of a 5$^3$ lattice for simulating the 79 charges of a Au nucleus) it is very
similar to what is observed in the nuclear experiment 
(Fig.~\ref{fig:zbound}). For specific values of the bond parameter 
in the transition region, the distributions of the 3-cluster asymmetry
$c_1-c_2-c_3$ (the cluster sizes $c_i$ are ordered in magnitude)
exhibit two bumps (Fig.~\ref{fig:perco2}). For the smaller 
lattice of 125 sites, the distribution extends over a major part of the 
interval [0,1] that is accessible after normalization with respect to 
the number of sites $c_0 = L^3$. Also this feature is reminiscent of the
result obtained for the $^{197}$Au fragmentation (Fig.~\ref{fig:spect}). 

\begin{figure} [!htb]	
    \leavevmode
    \centering
    \epsfysize=10.5cm

   \epsffile{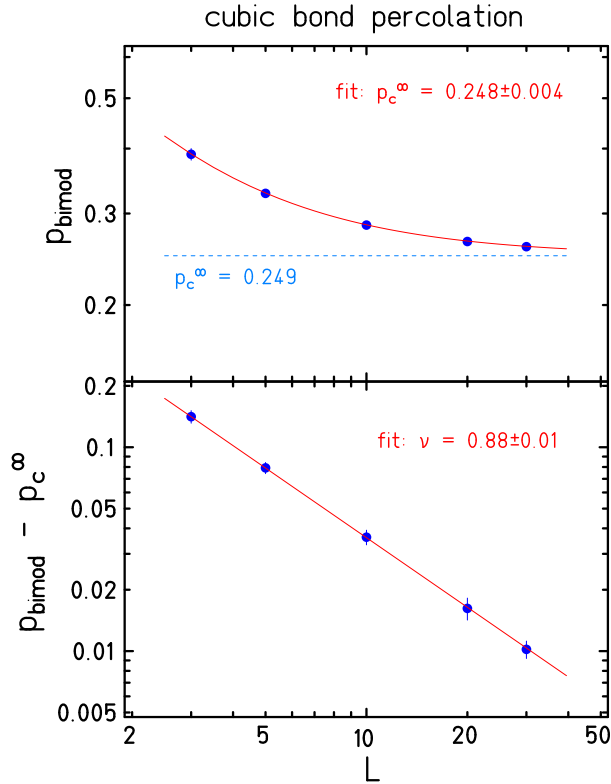}


\caption{{Bond parameter $p_{\rm bimod}$ for which the bimodal patterns 
appear
most clearly (top) and difference $p_{\rm bimod}-p_{c}^{\infty}$ (bottom) 
as a function of the lattice size $L$. 
The full lines represent the results of three-parameter (top) and 
two-parameter (bottom, with $p_{c}^{\infty}$ fixed) power-law fits 
according to Eq.~\protect\ref{eq:power}. The dashed line indicates the 
location of the critical bond parameter $p_{c}^{\infty}=0.249$ 
for the infinite system 
\protect\cite{stauffer,pbinf}.}}

\label{fig:power} 
\end{figure}

For the larger lattice, the distribution is still double-humped but
becomes much narrower. The bond parameter $p_{\rm bimod}$ at which the 
bimodal structure is most pronounced is smaller and much closer to the 
critical value. Calculations performed for 
various lattice sizes and samples of up to 200000 events
show that this variation is systematic and 
confirm that the law of finite-size scaling 
\cite{muell_borm} is obeyed by $p_{\rm bimod}$.
A power law fit according to the expression 

\begin{equation} \label{eq:power}
p_{\rm bimod}-p_{\rm c}^{\infty} = c\cdot L^{-1/{\nu}}
\end{equation}

\noindent
shows that the critical bond parameter for the infinite lattice is indeed 
approached by $p_{\rm bimod}$ (Fig.~\ref{fig:power}). 
A two-parameter fit with a fixed 
$p_{c}^{\infty}$~= 0.249 yields $\nu = 0.88 \pm 0.01$ in agreement with the 
known value $\nu = 0.88$ of the critical exponent describing the
divergence of the correlation length \cite{stauffer}.

\begin{figure} [!htb]	
    \leavevmode
    \centering
    \epsfxsize=0.65\linewidth

   \epsffile{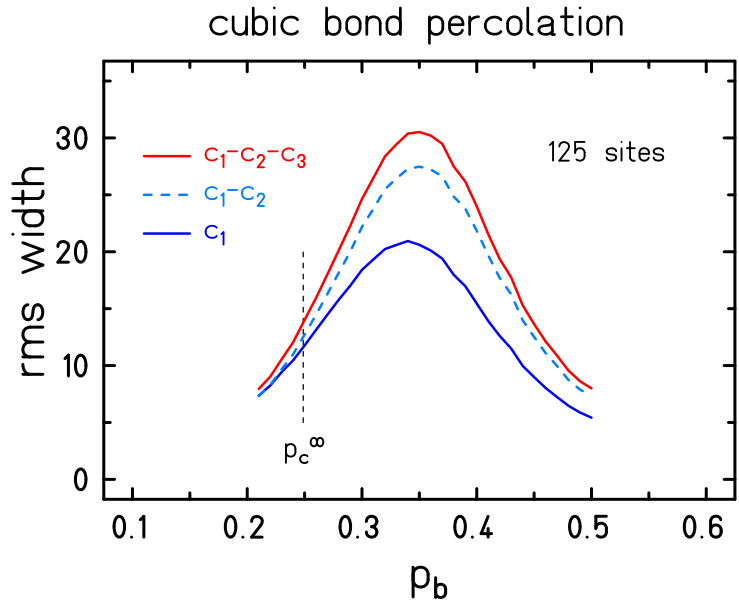}


\caption{{Root-mean-square widths of the $c_1$ (full line, blue), 
$c_1-c_2$ (dashed), and $c_1-c_2-c_3$ (full line, red) distributions 
as a function of the bond probability $p_b$.}}

\label{fig:cfluct} \end{figure}

\begin{figure} [!htb]	
    \leavevmode
    \centering
    \epsfxsize=0.65\linewidth

   \epsffile{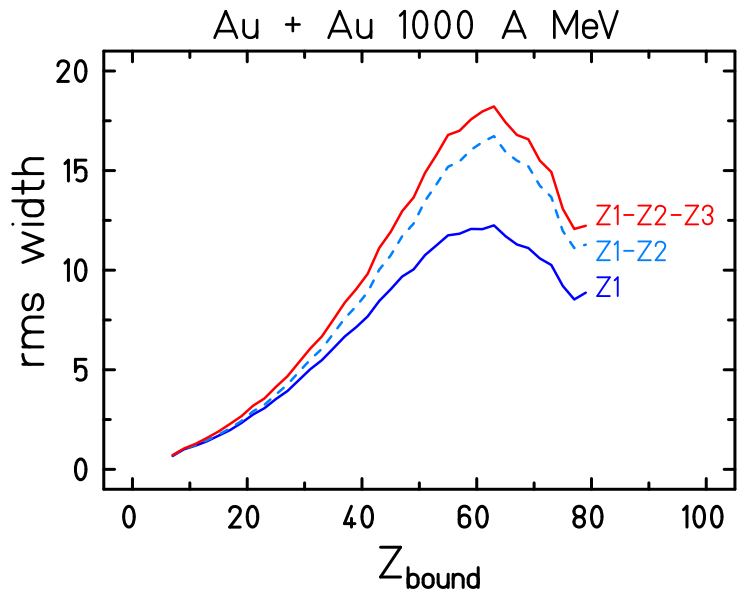}


\caption{{Root-mean-square widths of the $Z_1$ (full line, blue), 
$Z_1-Z_2$ (dashed), and $Z_1-Z_2-Z_3$ (full line, red) distributions 
as a function of $Z_{\rm bound}$. Normalization with respect to 
$Z_{\rm bound}$ will shift the maxima to $Z_{\rm bound} \approx 55$, i.e.
into the center of the transition region 
(cf.~Fig.~\protect\ref{fig:zbound}).}}

\label{fig:zfluct} \end{figure}

Finite-size scaling identifies the observed phenomenon as originating
from order-parameter fluctuations near the percolation phase transition.
The same law with the same exponent is also valid for the locations of
the maxima of the slopes of the $\langle c_{1}\rangle$ vs. $p_b$ transition
which practically coincide with $p_{\rm bimod}$ 
(cf. Fig.~\ref{fig:perco1}), or for the widths of the transition 
region~\cite{stauffer}. The 
fluctuations of the largest cluster size cause corresponding fluctuations
of the differences and asymmetries (Fig.~\ref{fig:cfluct}), a property
that identically appears in the fluctuation widths of the charge 
differences or asymmetries observed in the fragmentation of $^{197}$Au 
(Fig.~\ref{fig:zfluct}). The existence of two bumps in the event
distributions appears as a generic feature of fragmentation processes, 
including that modeled with percolation. 
The mere observation of this phenomenon 
can thus not be considered as providing 
evidence for a first-order phase transition.

\section{Reaction scenarios}
\label{sec:moldy}

In classical molecular dynamics, maximum size fluctuations define a 
critical percolation line (Kert\'esz line), or a critical 
percolation band in finite systems, in the temperature-density 
phase diagram~\cite{campi03}. The Kert\'esz line, known from studies of the
lattice-gas model \cite{kertesz,campi97}, extends from the thermodynamical
critical point into the supercritical region of higher density and 
temperature and is considered generic for simple 
fluids. Its identification requires appropriate algorithms for 
the recognition of clusters in the dense medium. 
Equilibrium cluster-size distributions along the critical band exhibit 
a power-law behaviour and bimodality~\cite{campi03,campi_priv}. 

It is a particular characteristic of the classical-molecular-dynamics 
model that the distributions of so-defined clusters do not 
significantly change as the systems are allowed to expand freely to a
breakup point beyond which clusters can be recognized in configuration space 
\cite{campi03}. Their properties acquired by originating
from a phase space location in the 
critical region will be reflected in the asymptotic
distributions. The reaction scenario suggested by these calculations  
thus links the observed percolation-like phenomena to a truly critical
behaviour of large systems. 
The applicability of the model to nuclear fragmentation can be 
tested by searching for predicted non-equilibrium phenomena at breakup 
\cite{campi03}. One of them, a considerable difference between 
the internal temperatures of the emerging fragments and that of the 
environment, is also a result of quantum-molecular dynamics 
(QMD \cite{zbiri07}). 
The recent analysis of multifragmentation following $^{197}$Au + $^{197}$Au 
collisions in the energy range 60A to 150A MeV with this model
has, in particular, also shown that bimodality is observed
and that the experimental asymmetry distributions of the largest fragments 
are reproduced rather well with QMD \cite{zbiri07}. 

On the other hand, phenomena resembling critical behaviour as it appears
in large systems 
are also observed for equilibrium distributions of small systems generated
within their coexistence zones. 
For the lattice-gas model, it has been rather generally shown that the 
observation of scaling inside the coexistence zone is 
compatible with a first-order phase transition because of finite 
size-effects \cite{gulm99}. The scaling will disappear in large systems.
Similar conclusions were reported by the authors of~\cite{pan98}.
When the Statistical Multifragmentation Model was used to describe
the fragmentation of relativistic $^{197}$Au projectiles, 
the experimentally observed power-law $Z$ and bimodal $Z_{\rm max}$ 
distributions in the transition region have been reproduced
with conditions below the critical point of this model \cite{botv95}.
These phenomena thus seem to appear naturally when viewing 
phase transitions in small systems through their partitioning into 
fragments. 

An interesting experimental observation is the coincidence of several 
signals considered indicative of a phase transition in fragmentation data
\cite{rivet_prag,borderie04}. Besides bimodality, 
this includes universal fluctuations 
($\Delta$-scaling \cite{botet01}) of the size of the largest fragment and 
the kinetic-energy fluctuations which have been associated with negative 
heat capacity \cite{dagostino00}. Searching for a common origin, it 
seems most likely that they are all related to the 
configurational fluctuations \cite{campi04,gulm06} 
to be expected in fragmentation processes 
and identified as finite-size effects in percolation. 

\section{Summary}
\label{sec:summ}

Experimental results regarding the largest fragment charge and the
asymmetries of the two and three largest fragments in the decays of 
$^{197}$Au projectile spectators at 1000 MeV per nucleon 
have been presented. The bimodal 
distributions at specific values of the sorting variable $Z_{\rm bound}$
reflect the size fluctuations of the largest fragment in the transition 
region between the regimes of residue production and of multifragmentation.
In the reaction scenario suggested 
by molecular dynamics, these configurational fluctuations are related
to a critical percolation region in the phase diagram which reduces to a
critical percolation line (Kert\'esz line) in large systems. 
In small systems, critical-like phenomena like scaling, power-law cluster 
distributions and bimodality are also exhibited by equilibrium distributions
generated at locations within the coexistence region. 
The underlying configurational fluctuations, identified as finite-size 
effects with percolation, thus seem generic for fragmentation processes
in small systems.
\\

Stimulating discussions with X. Campi and E. Plagnol are gratefully acknowledged.

\end{document}